\documentclass[10pt]{article}
\usepackage{epsf} 
\usepackage{amsmath}
\usepackage{amssymb}
\usepackage{epsfig}
\usepackage{latexsym}
\usepackage{amsfonts}
\usepackage{graphicx}%
\usepackage{varioref}
\usepackage{ifthen}
\begin{document}
\parindent 0mm 
\setlength{\parskip}{\baselineskip} 
\thispagestyle{empty}
\pagenumbering{arabic} 
\setcounter{page}{0}
\mbox{ }
\rightline{UCT-TP-275/08}
\newline
\rightline{August 2008}
\newline
\vspace{0.1cm}
\begin{center}
{\large {\bf 	Light quark masses from QCD sum rules with minimal hadronic bias}}
{\LARGE \footnote{{\LARGE {\footnotesize In Memoriam: Giuseppe (Beppe) Nardulli, Jan Stern, and Francisco (Paco) Yndurain, good friends and colleagues. Work supported in part by  NRF (South Africa) and DFG (Germany). Invited talk given by C.A. Dominguez at QCD-08, 14th International QCD Conference, Montpellier, France, July 2008.}}}}
\end{center}
\vspace{.05cm}
\begin{center}
{\bf C.A. Dominguez}$^{(a)}$,$^{(b)}$, {\bf N.F. Nasrallah}$^{(c)}$, {\bf R. R\"{o}ntsch}$^{(a)}$,\\
 {\bf K. Schilcher}$^{(d)}$
\end{center}
\begin{center}
$^{(a)}$Centre for Theoretical Physics and Astrophysics,
University of
Cape Town, Rondebosch 7700, South Africa\\
$^{(b)}$ Department of Physics, Stellenbosch University, Stellenbosch 7600, South Africa\\
$^{(c)}$ Faculty of Science, Lebanese University, Tripoli, Lebanon\\
$^{(d)}$ Institut f\"{u}r Physik, Johannes Gutenberg-Universit\"{a}t
Staudingerweg 7, D-55099 Mainz, Germany
\end{center}
\vspace{0.2cm}
\begin{center}
\textbf{Abstract}
\end{center}
\noindent
The light  quark masses are determined using a new QCD Finite Energy Sum Rule (FESR) in the pseudoscalar channel. This FESR involves an integration kernel designed to reduce considerably the contribution of the (unmeasured) hadronic resonance spectral functions. The QCD sector of the FESR includes perturbative QCD (PQCD) to five loop order, and the leading non-perturbative terms. In the hadronic sector the dominant contribution is from the pseudoscalar meson pole. Using Contour Improved Perturbation Theory (CIPT) the results for the quark masses at a scale of 2 GeV are $m_u(Q= 2\, \mbox{GeV}) = 2.9 \,\pm\, 0.2\,\mbox{MeV}$, $m_d(Q= 2\, \mbox{GeV}) = 5.3 \,\pm\,0.4 \, \mbox{MeV}$, and $m_s(Q= 2 \,\mbox{GeV}) = 102 \,\pm\,8 \, \mbox{MeV}$, for $\Lambda = 381\,\pm\, 16\, \mbox{MeV}$, corresponding to $\alpha_s(M_\tau^2) = 0.344 \,\pm0.009$. In this framework the systematic uncertainty in the quark masses from the unmeasured hadronic resonance spectral function amounts  to less than 2 - 3 \%. The remaining uncertainties above arise  from those in $\Lambda$, the unknown six-loop PQCD contribution, and the gluon condensate, which are all potentially subject to improvement.
\newpage
\noindent
The QCD running quark masses, together with the running coupling constant, are the most important parameters of the strong sector of the Standard Model of Particle Physics. Many attempts have been made over the years to extract the values of the quark masses from hadronic observables in various theoretical frameworks. This pursuit started with current algebra, and later chiral perturbation theory, which fixes light quark mass ratios \cite{RATIO}. This was followed by QCD sum rules \cite{OLD1}-\cite{OLD2}, and lattice QCD \cite{LATTICE}.
In the framework of QCD sum rules the ideal Green's function is the pseudoscalar correlator, as it involves the quark masses as an overall multiplicative factor (with subdominant and small quark mass corrections), and  the well known pseudoscalar meson pole. Unfortunately, the hadronic resonance spectral function is not directly known from experiment. In fact, except at best for the masses and widths of the first two radial excitations in the non-strange and strange sector, nothing is known about the actual shape of these spectral functions. This information is hardly enough to realistically reconstruct them; inelasticity and non-resonant background would remain unknown. We describe here a recent attempt \cite{COND}-\cite{QM} to reduce considerably this systematic uncertainty by incorporating a specific analytic kernel in Cauchy's theorem used to derive the FESR. This kernel takes the form of a second degree polynomial which vanishes at the peaks of the two radial excitations of the ground state pseudoscalar meson (pion or kaon). The idea of including polynomial kernels in FESR is not new; several different forms have been used in the past for various reasons and in various channels. What is new, though, is the requirement that the kernel vanishes at the peaks of the resonances entering the experimentally unknown spectral functions. This kernel does distort whatever model one uses as a guess for the resonance spectral function. But then, it also distorts the pseudoscalar meson pole, and the QCD contribution. Since Cauchy's theorem is still valid, so is the distorted FESR. Remarkably, this procedure not only  reduces considerably the importance of the unknown sector, but at the same time it leads to an exceptionally broad  region of stability against changes in $s_0$, the upper limit of integration in the FESR, which in CIPT is $s_0 \simeq 1 - 4 \, \mbox{GeV}^2$. We consider the correlator of (light) axial-vector current divergences

\begin{equation}
\psi_{5}(q^{2})|^j_i =
i \, (m_j + m_i) \,\int d^{4}x\; e^{i q x}
\times <T|(j_5(x),j_5^{\dagger}(0))|>|^j_i \;, 
\end{equation}

where $j_5(x)|^j_i = :\overline{q}_j(x) \,i \, \gamma_{5}\, q_i(x):\;$, and $i,j$ are quark flavours. To simplify the notation  we shall use in the sequel $m_j + m_i \equiv m$. We define the integration kernel

\begin{equation}
\Delta_5(s) = 1 - a_0 \;s - a_1\; s^2 \;,
\end{equation}

where $a_{0}$, and $a_1$ are free parameters to be fixed by the requirement that $\Delta_5(M_1^2) = \Delta_5(M_2^2) = 0$, with $M_{1,2}$ the masses of the two resonances in the  pseudoscalar channel, $\pi$(1460) and $\pi$(1830) for the non-strange channel, and K(1460) and K(1830) for the strange channel.
Invoking Cauchy's theorem for the second derivative of $\psi_5(q^2)$ one finds \cite{QM}

\begin{equation}
-\frac{1}{2\pi i}
\oint_{C(|s_0|)}
ds \psi_{5}^{'' QCD}(s)\,[F(s) - F(s_0)] = 2f_P^2
\, M_P^4 \Delta_5(M_P^2)+
\frac{1}{\pi} \; \int_{s_{th}}^{s_0}
ds \; Im \psi_{5}(s)|_{RES}\;\Delta_5(s),
\end{equation}

where

\begin{equation}
F(s) = - s \left(s_0 - a_0\,\frac{s_0^2}{2} - a_1\, \frac{s_0^3}{3} \right) + \frac{s^2}{2} 
- a_0\, \frac{s^3}{6} - a_1\, \frac{s^4}{12} \;.
\end{equation}

\begin{figure}
\begin{center}
\includegraphics[height=2.7 in,width=4 in]
{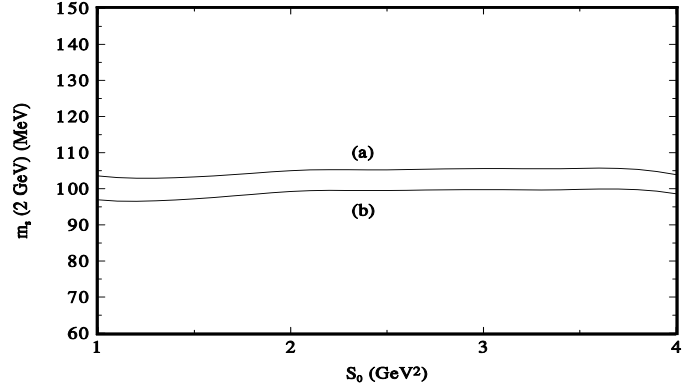}
\caption{Strange quark mass at 2 GeV for $\Lambda_{QCD} = 365 \;(397)\; \mbox{MeV}$, curves (a) and (b), respectively.}
\end{center}
\end{figure}

The function $\psi^{'' QCD}_5(q^2)$ is currently known up to five-loop order in PQCD \cite{PQCD5}. We define the left hand side of Eq.(3) as\

\begin{equation}
\delta_5(s_0)|_{QCD} \equiv - \frac{1}{2\pi i}
\oint_{C(|s_0|)}
ds \psi_{5}^{''}(s)|_{QCD} 
\; [F(s) - F(s_0)] , 
\end{equation}

and obtain in PQCD and in the framework of CIPT \cite{QM}

\begin{equation}
\delta_5(s_0)|_{PQCD} = \frac{\overline{m}^2(s_0)}{16 \pi^2} \,\sum_{j=0}^4 K_j 
\frac{1}{2 \pi} 
  \int_{-\pi}^{\pi}dx  
   \Big[ F(x)- F(s_0) \Big]   [a_s(x)]^j 
exp \Bigg[ - 2 i \sum_{M=0} \gamma_M 
\int_0^x dx' [a_s(x')]^{M+1} \Bigg]  \;,
\end{equation}

where the RGE for the mass and coupling have been used, $a_s(x) \equiv \alpha_s(x)/\pi$, and where 

\begin{equation}
F(x) = \sum_{N=1}^4 (-)^N \; b_N \; s_0^N \; e^{iNx} \;,
\end{equation}

with the angle $x$   defined through $s = s_0 \, e^{ix}$, and  $x \in (- \pi, \pi)$.  

The constants above are ($n_F = 3$): $K_0= C_{01}$, $K_1 = C_{11} + 2 C_{12}$, $K_2 = C_{21} + 2 C_{22}$, $K_3 = C_{31} + 2 C_{32}$, $K_4 = C_{41} + 2 C_{42}$, with 
$C_{01} = 6$, $C_{11} = 34$, $C_{12} = - 6$, $C_{21} = - 105 \;\zeta(3) + 9631/24$, 
$C_{22} = - 95$, $C_{23} = 17/2$, $C_{31}= 4748953/864 - \pi^4/6 - 91519\; \zeta(3)/36 + 715 \;\zeta(5)/2$,
$C_{32} = - 6 \;[4781/18 - 475 \;\zeta(3)/8]$, $C_{33} = 229$, $C_{34} = - 221/16$, $C_{41} = 33 532.26$, $C_{42} = - 15 230. 6451$, $C_{43} = 3962.45493$, $C_{44} = - 534.052083$, $C_{45} = 24.1718750$, and $\zeta(x)$ is Riemann's zeta function. 
Finally, $b_1= -(s_0 - a_0 s_0^2/2 - a_1 s_0^3/3)$, $b_2 = 1/2$, $b_3 = - a_0/6$, and $b_4 = - a_1/12$.
Regarding the value of $\Lambda_{QCD}$ entering $\alpha_s(s_0)$,  it can be extracted from the strong coupling obtained from $\tau$-decay . We use the latest high precision result of \cite{ALEPH2}, i.e. $\alpha_s(M_\tau^2) = 0.344 \,\pm0.009$, leading to $\Lambda_{QCD} = 365 - 397$ MeV.\\

The light-quark condensate contribution to the left hand side of Eq.(3) is

\begin{eqnarray}
\delta_5(s_0)|_{<\bar{q} q>} &=& - 2 \;\;\frac{\overline{m}^2(s_0)}{s_0^2} \; \;\langle m_s \overline{q} q \rangle|_{\mu_0}\;\;
\frac{1}{2 \pi} \int_{-\pi}^\pi dx \;e^{-2 i x} 
\Big[F(x) - F(s_0)\Big] \Big[ 1 + \frac{23}{3} a_s(x)\Big] \nonumber \\ [.3cm]
&\times& exp  \Bigg[ - 2 i \sum_{M=0} \gamma_{M}  \int_0^x dx' [a_s(x')]^{M+1} \Bigg] \;,
\end{eqnarray}

\begin{figure}[ht]
\begin{center}
\includegraphics[height=2.5 in,width=3.0 in]{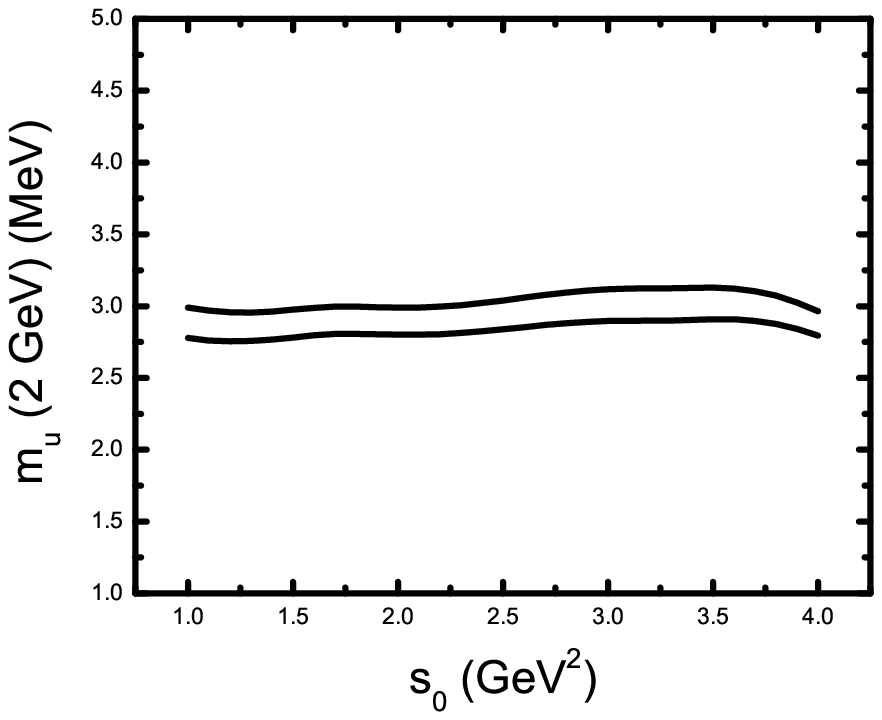}
\caption{Up quark mass at 2 GeV for $\Lambda_{QCD} = 365 \;(397)\; \mbox{MeV}$, upper (lower) curve, respectively.}
\end{center}
\end{figure}
\begin{figure}
\begin{center}
\includegraphics[height=2.5 in,width=3 in]
{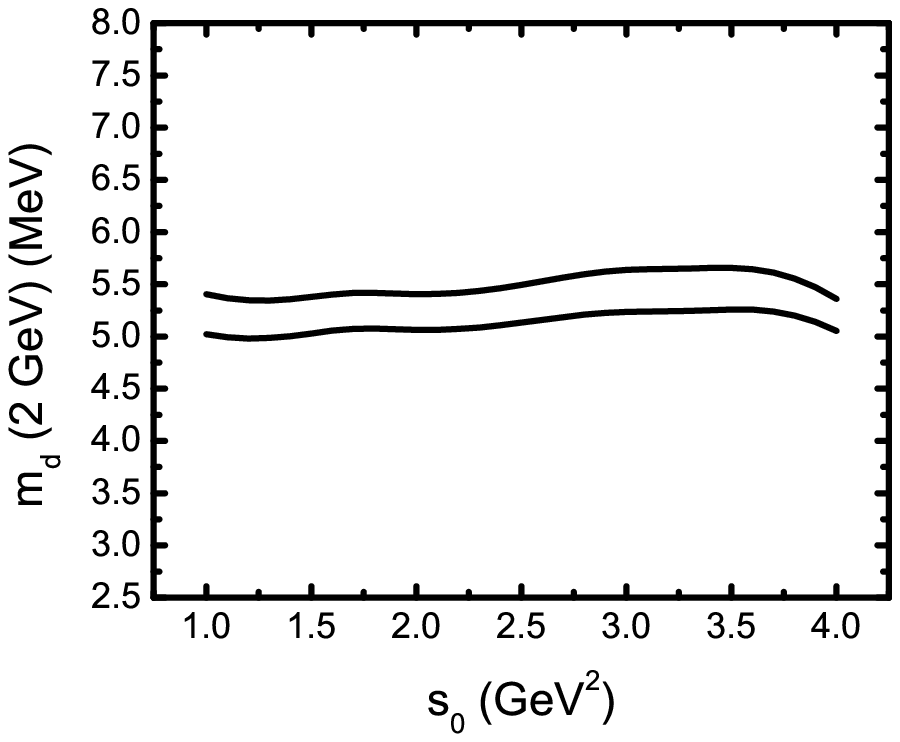}
\caption{Down quark mass at 2 GeV for $\Lambda_{QCD} = 365 \;(397)\; \mbox{MeV}$, upper (lower) curve, respectively.}
\end{center}
\end{figure}

and the contribution of the gluon condensate becomes

\begin{eqnarray}
\delta_5(s_0)|_{<G^2>} &=& \frac{1}{4} \;\;\frac{\overline{m}^2(s_0)}{s_0^2} \; \;\langle \frac{\alpha_s}{\pi} G^2\rangle|_{\mu_0}\;
\frac{1}{2\pi} \int_{-\pi}^\pi dx \;e^{-2 i x}
 \Big[F(x) - F(s_0)\Big] 
  \Big[ 1 + \frac{16}{9} a_s(\mu_0)+ \frac{121}{18} a_s(x)\Big] 
 \nonumber \\ [0.3cm]
  &\times& exp  \Bigg[ - 2 i \sum_{M=0} \gamma_M  \int_0^x dx' [a_s(x')]^{M+1} \Bigg] \;,
\end{eqnarray}

where the scale $\mu_0 \simeq 1 \; \mbox{GeV}^2$ appears in connection with the removal of logarithmic quark mass singularities (see \cite{OLD2}). The  dimension six four-quark condensate, and the higher order quark mass corrections turn out to be negligible in this application. We shall use  $\langle \frac{\alpha_s}{\pi} G^2 \rangle \simeq 0.06\; \mbox{GeV}^4$, and $\langle \bar{q}\,q \rangle \simeq ( - 250 \;\mbox{MeV})^3$ \cite{CADTAU}.
CIPT \cite{CIPT} has been shown to provide better convergence than Fixed OPT in the QCD analysis of the vector and axial-vector correlators in tau-lepton decay \cite{CADTAU}-\cite{CIPT}. We found this to be also the case in the determination of the light quark masses \cite{QM} . Unlike the case of FOPT, where $\alpha_s(s_0)$ is frozen in Cauchy's contour integral and the RG is implemented after integration, in CIPT $\alpha_s$ and the quark mass are running and the RG is used before integrating. This is done  through a single-step numerical contour integration and using as input the strong coupling and the quark mass obtained  by solving numerically the RG equation. Notice that the invariant quark mass enters the expression of the running quark mass as an overall multiplicative factor; this invariant mass is what one determines from the FESR.
Turning to the hadronic resonance sector, we have used a parametrization involving two Breit-Wigner forms normalized at threshold according to chiral perturbation theory, as first proposed in \cite{THR}. It is unnecessary to improve on this parametrization as the presence of the integration kernel in the FESR reduces this contribution  to the level of 2 - 3 \%. In Figs. (1)-(3) we show the results for the strange, the up, and the down quark masses at $Q = 2 \,\mbox{GeV}$. The region of stability, $s_0 \simeq 1\,-\, 4 \,\mbox{GeV}^2$, is exceptionally wide, with a remarkably low  spread of 4 - 5 \% for the up and down quark masses, and 1 - 2 \% for the strange quark mass. To achieve a reasonable error we have allowed for a $ \pm 30\, \%$  uncertainty in the hadronic resonance parametrizations,  a factor two uncertainty in the value of the gluon condensate, and have assumed the (unknown) six-loop contribution to be equal to the five-loop one. In order to disentangle the strange quark mass from the sum $(m_s + m_q)$ (q=u,d) we have used the chiral perturbation theory ratio \cite{RATIO} $m_s/m_{ud} = 24.4 \,\pm\,1.5$, with $m_{ud} \equiv (m_u+m_d)/2$. Similarly, to disentangle e.g. the up quark mass we used $m_u/m_d = 0.553 \,\pm\,0.043$. This leads to the values
\begin{equation}
m_u (2\,\mbox{GeV}) = 2.9\,\pm\,0.2\,\mbox{MeV}\;, 
\end{equation}

\begin{equation}
m_d (2\,\mbox{GeV}) = 5.3\,\pm\,0.4\,\mbox{MeV}\;,
\end{equation}

\begin{equation}
m_s (2\,\mbox{GeV}) = 102\,\pm\,8\,\mbox{MeV}\;,
\end{equation}

\begin{equation}
\frac{m_u + m_d}{2}  = 4.1\,\pm\,0.2\,\mbox{MeV}\;,
\end{equation}

\begin{equation}
\frac{m_d + m_u}{m_d - m_u}  = 0.29\,\pm\,0.05\;.
\end{equation}

These results are in agreement with some recent lattice determinations \cite{LATTICE}.\\
 
This work has been supported by NRF (South Africa) and DFG (Germany).

\end{document}